# COMPUTAÇÃO: O VETOR DE TRANSFORMAÇÃO DA SOCIEDADE

Avelino Francisco Zorzo[1], André Luís Alice Raabe[2], Christian Brackmann[3]

## INTRODUÇÃO

A sociedade está mudando, sempre mudou e continuará mudando. Entretanto, as mudanças estão cada vez mais rápidas; o que antes ocorria de uma geração para a próxima agora ocorre dentro de uma mesma geração. A Computação tem sido um dos vetores do aumento da velocidade destas mudanças e é uma área que permeia, atualmente, todas as demais áreas do conhecimento.

Os conhecimentos em Computação são tão importantes para a vida na sociedade contemporânea quanto os conhecimentos básicos de Matemática, Filosofia, Física, dentre outras, assim como contar, abstrair, pensar, relacionar, ou medir. Desta forma, na sociedade atual e futura é fundamental que todos os indivíduos tenham conhecimentos básicos de Computação.

Por um lado, o uso de tecnologia tem se tornado cada vez mais presente. Por exemplo, atualmente uma pessoa pode ler um livro utilizando um leitor digital (*tablet*, *kindle* ou similar); quando falamos ao telefone, podemos visualizar a outra pessoa por vídeo; processos de cooperação para desenvolvimento de novas tecnologias ou novas pesquisas podem ser feitas de maneira instantânea com pessoas do mundo todo. Entre os diversos fatores que tem auxiliado nestas mudanças podemos citar o aumento no número de pessoas que estão conectadas à Internet, a expansão da telefonia celular para os mais diversos lugares do mundo, ou o aumento na geração e compartilhamento de dados. O poder de processamento dos computadores segue aumentando, e é possível que em poucos anos se equipare em alguns aspectos ao poder de processamento do cérebro humano.

O impacto da Computação nas outras áreas do conhecimento também é cada vez maior e mais profundo. Problemas complexos de diferentes áreas da ciência estão agora sendo abordados com uma perspectiva computacional, uma vez que a Computação provê estratégias e artefatos para lidar com a complexidade, avançando na solução de problemas que há poucos anos não seriam possíveis.

Exemplos notórios podem ser encontrados na Biologia, no mapeamento do genoma humano, na identificação de variações de enzimas, na simulação da adaptação de seres vivos em diferentes ambientes. Na Saúde, a Computação atua no desenvolvimento de medicamentos, realização de cirurgias remotas e até mesmo na simulação de previsão de tempo de contaminação por uma doença em um determinado ambiente. Na Química a Computação possibilita simular reações químicas sem a necessidade de expor pesquisadores a situações de risco de vida. A Estatística é uma área da Matemática que impacta fortemente no fazer científico e que foi amplamente alterada a partir da abordagem Computacional. Os exemplos se estendem para outras áreas como Arquitetura, Agronomia, Direito, Economia, Educação, Engenharia, Física, Psicologia, Segurança ou Zoologia. Enfim é difícil encontrar uma área do conhecimento que não esteja sendo impactada e até transformada pela abordagem computacional.

A quantia de dados existentes combinada com as possibilidades atuais de processamento computacional podem ajudar a melhorar a vida das pessoas quando forem utilizados para a criação de soluções inovadoras. Por exemplo, dados sobre crimes que acontecem em diversos lugares de uma cidade podem ser analisados e processados para estimar onde os crimes podem acontecer, economizando assim recursos. Ou ainda, armazenar informações sobre alimentos (agricultura, pecuária, pesca, ...) em um local único de forma a gerenciar o suprimento de comida de maneira global, podendo fazer, assim, uma melhor distribuição de alimentos.

---

[1] Diretor de Educação da Sociedade Brasileira de Computação - SBC. Professor Titular da PUCRS. *avelino.zorzo@pucrs.br*

[2] Vice-coordenador da Comissão Especial em Informática na Educação da SBC. Professor da UNIVALI. *andré.raabe@univali.br*

[3] Professor do Instituto Federal Farroupilha. Doutorando em Informática na Educação na UFRGS. *brackmann@iffarroupilha.edu.br*

A forma como a tecnologia está sendo utilizada pelas pessoas, ou mesmo a diminuição no tamanho e no preço dos equipamentos, nos permite levá-la para locais onde há poucos anos não era possível. Por exemplo, um telefone celular com uma câmera pode ser transformado em um microscópio para diagnosticar doenças em locais onde não existe uma infraestrutura médica adequada. Carros e semáforos podem "falar" entre si, de forma a melhorar as condições de tráfego, ou para facilitar o deslocamento de uma ambulância até um acidente ou até o hospital. Se avançarmos mais uns 30 ou 40 anos, com o aumento do poder de processamento dos computadores poderemos ter um cenário cujos limites ainda não são conhecidos.

Para que este potencial de mudança se concretize é necessário formar cidadãos aptos não apenas a lidarem com a tecnologia como usuários, mas sim a conceberem e produzirem tecnologia. É neste ponto que o conhecimento em Computação passa a ser o grande diferencial. As pessoas precisam resolver problemas de maneira flexível ou adaptável, estejam elas imaginando novas soluções, desenvolvendo novos equipamentos ou o software que irá ser utilizado nestes equipamentos. Elas precisam pensar além "do que é" para "o que poderia ser".

Neste sentido questiona-se: Quais habilidades os estudantes precisam para se preparar para este futuro? A resposta tem sido construída pela comunidade acadêmica de diversos países e uma destas habilidades converge para uma expressão denominada "Pensamento Computacional". Podemos rapidamente definir Pensamento Computacional como sendo "a combinação do pensamento crítico com os fundamentos da Computação, criando uma metodologia para resolver problemas" (WING, 2006). Muitas iniciativas de uso de Computação nas escolas já possuem elementos do Pensamento Computacional, entretanto o esforço da comunidade acadêmica da área possibilitou a construção de um vocabulário comum e um modelo para qualificar e direcionar os esforços dos professores e gestores educacionais.

Desta forma, a noção sobre Pensamento Computacional será aprofundada nas próximas seções deste artigo, ilustrando também as principais pesquisas e iniciativas que vem sendo realizadas no mundo e Brasil ao longo dos anos na área. Nas conclusões argumentaremos sobre a importância de incluir a Computação como um dos fundamentos para a Educação Básica. Mas gostaríamos antes de fazer uma distinção fundamental entre o ensino de Computação ou Pensamento Computacional e de Informática ou Tecnologias da Informação e Comunicação (TIC).

Historicamente, as políticas públicas de fomento às TIC na educação – como é o caso do Programa Nacional de Tecnologia Educacional (ProInfo), do governo federal – deram maior destaque para a implantação de infraestrutura tecnológica nas escolas. Mais recentemente começaram a se fortalecer estratégias de capacitação do professor para o uso dessas ferramentas tecnológicas como instrumento pedagógico e para a produção de conteúdos digitais. Em 2014, 85% das escolas Brasileiras possuíam laboratório de computadores e 92% possuíam alguma forma de conexão com a Internet. As atividades realizadas pelos professores e estudantes com estes equipamentos em sua maioria estão associadas à pesquisa e produção de conteúdos e em poucos casos o uso de software educacional e de objetos de aprendizagem (TIC Educação, 2015).

A utilização da Informática nestes moldes pode favorecer a ilustração de conceitos e a aprendizagem, no entanto, em maior profundidade não exploram plenamente o potencial de desenvolvimento cognitivo dos estudantes. Não os habilitam a se tornarem criadores de tecnologia. O ensino de fundamentos e conceitos de Computação podem dotar o estudante de ferramentas mentais e artefatos para sistematizar a solução de problemas e construir modelos computacionais tais como programas e simulações, e com isso tornarem-se protagonistas das mudanças na sociedade tecnológica. Ainda, na mesma direção, as iniciativas de ensino de programação de computadores nas escolas podem não ser suficientes para alcançar estes objetivos, pois carecem de um tratamento mais científico e direcionado para resolução ampla de problemas.

Entendemos que fomentar a aplicação do Pensamento Computacional na Educação Básica é o melhor caminho para promover a transformação da sociedade de forma crítica e conectada com os valores e com a cultura nacional.

**PENSAMENTO COMPUTACIONAL**

Na década de 1940, John von Neumann profetizou que computadores não seriam apenas ferramentas para ajudar a ciência, mas também uma forma de fazer ciência. Entre as décadas de 1950 e 1960 surgiu o termo "Pensamento Algorítmico" que era compreendido como "orientação mental para formular problemas com conversões, com alguma entrada (*input*) para uma saída (*output*), utilizando uma forma algorítmica para executar as conversões" (DENNING, 2009).

Algumas décadas depois, no ano de 1975, o ganhador do Prêmio Nobel de Física, Laureate Ken Wilson promoveu a ideia que a Simulação e a Computação eram uma forma de fazer ciência que não estavam disponíveis anteriormente. Foi através do uso das máquinas que o ganhador do prêmio Nobel conseguiu criar modelos computacionais que produziram uma compreensão nunca antes imaginada sobre a mudança de estados de materiais. No início da década seguinte, Laureate se uniu a outros cientistas de destaque para tentar defender que a Computação seria a solução para os grandes desafios da ciência. A partir dos levantamentos realizados pelos pesquisadores, definiu-se que a Computação seria o terceiro pilar da ciência, além dos tradicionais: Teoria e Experimentação. Nasce a base para o que mais tarde viria a ser chamado de "Pensamento Computacional".

Papert (1980, p. 182) foi o primeiro autor utilizar o termo "Pensamento Computacional" na literatura, mais especificamente em seu livro que trata da cultura dos computadores e o papel da tecnologia no ensino de crianças. Porém a sociedade ainda não estava apta a compreender naquele momento o potencial das idéias de Papert.

Somente em 2006 o termo Pensamento Computacional tornou-se amplamente conhecido a partir da publicação de um artigo por Wing (2006). A partir deste artigo o Pensamento Computacional passou a ser amplamente discutido. A ideia mais impactante do artigo é que o Pensamento Computacional é uma habilidade fundamental para qualquer um, não apenas para cientistas da Computação. E que juntamente com a leitura, a escrita e a aritmética, devíamos desenvolver o Pensamento Computacional como habilidade analítica em cada criança.

A popularidade obtida pela publicação de certa forma reflete a ânsia que existia na comunidade de pesquisadores de Educação em Computação, Tecnologia Educacional, Informática na Educação e áreas afins pela conquista de espaço e de argumentos que pudessem renovar e aprofundar o uso de tecnologia nas escolas.

Dentre as muitas definições de Pensamento Computacional que surgiram após o termo ter sido disseminado pelo artigo de Wing, a definição construída pela International Society for Technology in Education (ISTE), em conjunto com a Computer Science Teachers Association (CSTA), é a que fornece uma visão, ao mesmo tempo, objetiva e abrangente do termo. Nesta definição, Pensamento computacional é um processo de resolução de problemas que inclui (não somente) as seguintes características: (i) Formulação de problemas de forma que computadores e outras ferramentas possam ajudar a resolvê-los; (ii) Organização lógica e análise de dados; (iii) Representação de dados através de abstrações como modelos e simulações; (iv) Automatização de soluções através do pensamento algorítmico; (v) Identificação, análise e implementação de soluções visando a combinação mais eficiente e eficaz de etapas e recursos (vi) Generalização e transferência de soluções para uma ampla gama de problemas (CSTA, 2015).

Papert (2008) menciona que frequentemente as pessoas temem que ao usar modelos computacionais de pensamento poderá se levar a um pensamento mecânico ou linear. Entretanto, ele faz um contraponto ao defender o "pensar como um computador" como sendo mais uma ferramenta mental a ser utilizada, pois não leva os indivíduos a sempre pensarem da mesma forma, mas sim a estarem providos de mais uma forma de pensar ao se deparar com um problema.

A promessa do Pensamento Computacional é empoderar estudantes com as habilidades que eles precisam para se tornarem efetivos e confiantes solucionadores de problemas em um mundo complexo [...] com habilidades de pensamento computacional, estudantes irão reconhecer quando um computador pode ajudar a resolver um problema (CSTA, 2015, p. 05). Conforme Gonçalves (2015), não é pretendido que os indivíduos fiquem restritos a este tipo de pensamento, mas que, quando se confrontarem com uma situação complexa possam refletir e verificar se o Pensamento Computacional poderá ajudar a solucionar determinado problema.

**INICIATIVAS DE PROMOÇÃO DO PENSAMENTO COMPUTACIONAL NO MUNDO**

O fomento ao ensino de Computação e o Pensamento Computacional está na pauta das mudanças sociais em diversos países. Além de iniciativas governamentais, muitas empresas, organizações e universidades de alcance global têm produzido materiais, ferramentas e guias voltados a auxiliar na introdução do Pensamento Computacional na Educação (RAABE *et al.* 2015).

A seguir apresenta-se um panorama de como noções de Computação e Pensamento Computacional têm se relacionado com a Educação Básica em diversos países (lista em ordem alfabética e não exaustiva):

- Alemanha: desde 2008 adotou novos padrões que claramente distinguem o ensino de TICs e Ciência da Computação desde a segunda metade do Ensino Fundamental (JONES, 2011) (EPA, 2004);
- Argentina: em 2015 foi assinada a resolução que estabelece que "o ensino e a aprendizagem de 'Programação' é de importância estratégica no Sistema Educativo Nacional durante a Educação Básica para fortalecer o desenvolvimento econômico e social da Nação". Esta resolução cria também a "rede de escolas que programam" com objetivo de troca de experiências entre as escolas que ensinam programação (CFE Argentina, 2015);
- Austrália: no ano de 2015 houve a proposição de uma reestruturação dos currículo da Educação Básica onde a programação foi colocada como uma das principais competências, tornando a disciplina compulsória. A importância dada à Computação é tal que disciplinas como História e Geografia foram colocadas como facultativas (DAVIS, 2015) (DET, 2015);
- Coréia do Sul: o ensino da Ciência da Computação ocorre desde o ano 1987. No ano de 2018 entrará em vigor um novo currículo reforçando ainda mais a presença do Pensamento Computacional nas salas de aula (CHOI, 2015);
- Escócia: Desde 1980 o ensino de noções de Computação ocorre com estudantes a partir dos 14 anos de idade. Em 2011 houve uma reformulação do documento que rege o sistema de ensino do país dos 3 aos 18 anos, onde a Computação consta antes mesmo do Ensino Fundamental (SADOSKY, 2013);
- Estados Unidos da América: o país assinou no dia 10 de dezembro de 2015 a Lei Federal "Every Student Succeeds Act" (ESSA), responsável pelas políticas públicas do país. Nesse documento são detalhados, desde a forma como ocorrem os financiamentos, até a maneira como as escolas são avaliadas. A lei também coloca a Ciência Computação em condições de igualdade com outras disciplinas acadêmicas, tais como Matemática, Geografia, História, Inglês e Ciências. O documento não define como a implantação deve ocorrer, porém incentiva sua adoção e permite a obtenção de recursos para tal (âmbito federal e estadual). Atualmente, a maioria dos 50 estados do país já adotaram o ensino do Pensamento Computacional nas escolas. O grande propulsor do Pensamento Computacional no país foi através da ONG Code.Org, dedicada na expansão do acesso a Ciência da Computação que teve uma aceitação muito grande nos EUA, chamando a atenção de grandes empresas na área de TI e recebendo apoio das mesmas pela promoção de seus propósitos (GUZDIAL, 2014) (CSTA, 2015) (CODE.ORG, 2015) (USCONGRESS, 2015).
- Finlândia: a partir de 2016 adotará a Ciência da Computação desde o Ensino Fundamental de forma compulsória. A demanda surgiu da iniciativa privada para atender a demanda de profissionais com formação considerada adequada (MYKKÄNEN e LIUKAS, 2015) (WEINBERG, 2015).
- França: encontra-se a partir do ano de 2015 em processo de adoção de aulas de programação como atividades extracurriculares, incluindo disciplinas como: Fundamentos das Linguagens de Programação e Desenvolvimento de Aplicativos com a utilização de algoritmos simplificados (JOHNSON, 2015) (FLEURY, 2015);

- <u>Grécia</u>: o Pensamento Computacional é ensinado a partir do terceiro ano do Ensino Fundamental. A partir dos 10 anos, crianças já realizam atividades de programação (BALANSKAT e ENGELHARDT, 2014);
- <u>Índia</u>: oferece Ciência da Computação para alunos desde os 12 anos de idade. Algumas das disciplinas são: Programação, Redes de Computadores, Arquitetura de Computadores, etc. Um novo currículo que está sendo desenvolvido e deve integrar ainda mais elementos da Computação (SADOSKY, 2013);
- <u>Reino Unido</u>: o ensino da Ciência da Computação é considerado obrigatório em todos os quatro "Key-Stages" (Ensino Fundamental). No Ensino Médio, as disciplinas voltadas para programação são opcionais. Uma pesquisa recente aponta que 60% dos pais e 75% dos alunos preferem aulas da linguagem de programação Python ao invés do ensino de outro idioma no Ensino Fundamental (SCC, 2015) (TJL, 2015);
- <u>União Europeia</u>: a ONG "European Schoolnet" está tomando a frente da introdução do Pensamento Computacional nos currículos de 31 países europeus. Em 2015 alguns países-membros adotaram o Pensamento Computacional em seu currículo, entre eles: Áustria, Bélgica, Bulgária, República Tcheca, Dinamarca, Estônia, Finlândia, França, Hungria, Irlanda, Israel, Lituânia, Malta, Polônia, Portugal, Eslováquia, Espanha e Reino Unido (BALANSKAT e ENGELHARDT, 2014);

**INICIATIVAS DE PROMOÇÃO DO PENSAMENTO COMPUTACIONAL NO BRASIL**

No Brasil, até o momento da elaboração deste artigo, as politicas educacionais relacionadas a tecnologia estão restritas à abordagem de letramento e inclusão digital. Nenhum documento oficial menciona ensino de Fundamentos de Computação.

Em 2015 iniciou-se a construção da Base Nacional Curricular Comum (BNCC), a qual define os conhecimentos essenciais aos quais todos os estudantes brasileiros têm o direito de ter acesso e se apropriar durante sua trajetória na Educação Básica (BNCC, 2015). A versão inicial da BNCC não faz referência à área de Computação, mas apresentava Tecnologias Digitais como *tema integrador*. Dentro deste tema, a área de Linguagens é a que possui o maior número de referências, pois possuia um eixo em Língua Portuguesa chamado *Práticas Culturais das Tecnologias de Informação e Comunicação*. Ou seja, ainda não há no Brasil o reconhecimento da importância dada aos conhecimentos ligados a Computação da forma como ocorre em outros países. Nesse sentido, a Sociedade Brasileira de Computação (SBC) se articulou para solicitar modificações no teor da BNCC visando considerar a Computação como área de conhecimento (SBC, 2016). O resultado ainda não é conhecido.

Em contrapartida um corpo sólido de pesquisas e projetos envolvendo o ensino de Computação na Educação Básica são realizadas no Brasil desde a década de 1980. As iniciativas são muitas e bastante diversificadas. Em meados da década de 1980, Papert (1985) inicia o uso da linguagem Logo em escolas em todo mundo. No Brasil até o ano de 1996 muitos projetos foram realizados com programação dessa linguagem (Valente, 1996). O uso de robótica educacional que iniciou timidamente com kits de empresas como a Lego, hoje está amplamente disseminado em muitas escolas e instituições educacionais utilizando, inclusive, alternativas de baixo custo, envolvendo por vezes a reciclagem de componentes eletrônicos.

Diversas iniciativas de introdução ao Pensamento Computacional têm sido realizadas nos últimos anos envolvendo pesquisadores de escolas e universidades em diferentes níveis da educação escolar (BARCELOS E SILVEIRA, 2012; ANDRADE, 2013; FRANÇA e AMARAL, 2013; RIBEIRO *et al.*, 2013; VIEL, RAABE e ZEFERINO, 2014). O tema do Pensamento Computacional tem se tornado foco de muitos trabalhos de mestrado e doutorado cujos resultados são geralmente divulgados em conferências como o Workshop sobre Educação em Computação (no Congresso Anual da Sociedade Brasileira de Computação) e o Congresso Brasileiro de Informática na Educação.

Existem também esforços feitos pela Sociedade Brasileira de Computação (SBC) para a disseminação do Pensamento Computacional na Educação Básica no Brasil. Um exemplo é a Olimpíada Brasileira de Informática, "uma competição organizada nos moldes das outras olimpíadas científicas brasileiras, como Matemática, Física e Astronomia. O objetivo da OBI é despertar nos alunos o interesse por uma ciência importante na formação básica hoje em dia (no caso, Ciência da Computação), através de uma atividade que envolve desafio, engenhosidade e uma saudável dose de competição" (olimpiada.ic.unicamp.br).

Outras iniciativas merecedoras de registro são os CodeClubs[4] organizados por voluntários para levar atividades de programação para escolas, a criação de ambientes de programação em português como Portugol Studio (NOSCHANG, *et al.* 2014) e iniciativas de incentivo ao ensino de programação em larga escala como o Programaê[5] que tem cada dia alcançado mais adeptos.

Atualmente existem mais de 100 cursos de Licenciatura em Computação no Brasil. Estes cursos, segundo o Parecer 136/2012 do CNE/MEC, têm como objetivo principal "preparar professores para formar cidadãos com competências e habilidades necessárias para conviver e, prosperar em um mundo cada vez mais tecnológico e global e que contribuam para promover o desenvolvimento econômico e social de nosso País. A introdução do pensamento computacional e algorítmico na educação básica fornece os recursos cognitivos necessários para a resolução de problemas, transversal a todas as áreas do conhecimento. As ferramentas de educação assistida por computador e os sistemas de educação à distância tornam a interação ensino- aprendizagem prazerosa, autônoma e efetiva, pois introduzem princípios e conceitos pedagógicos na interação humano-computador. Essas ferramentas são desenvolvidas com a participação de Licenciados em Computação. Genericamente, todo sistema computacional com funcionalidade pedagógica ou que necessita de assistência para seu uso, requer a participação dos Licenciados em Computação".

**CONCLUSÕES**

Ciência da Computação é uma ciência dotada de mecanismos lógicos, linguísticos e tecnológicos para "resolução de problemas". Um desses mecanismos é conhecido pelo nome de "algoritmo". Esses problemas podem ser de qualquer natureza, por exemplo, no contexto de Administração, Antropologia, Biologia, Física, Direito ou Matemática; enfim, problemas do cotidiano.

O processo cognitivo usado pelos seres humanos para resolver problemas por meio de algoritmos, como já mencionado anteriormente, é chamado de Pensamento Computacional. Este processo, diferente de raciocínio lógico e matemático, habilita os estudantes a especificar e organizar a solução de problemas a partir do desenvolvimento de habilidades como abstração, refinamento, modularização, recursão, metacognição, entre outras. Habilidades estas que, ao se tornarem parte do repertório cognitivo dos indivíduos, impactam na sua forma de relacionar com o mundo.

O domínio da Computação e das Tecnologias da Informação e Comunicação é fundamental e estratégico para o desenvolvimento social e econômico de uma nação. Esse domínio fundamenta-se em um fluxo contínuo de aprendizado, disseminação e evolução do conhecimento e tecnologias subjacentes, com diversos atores: estudantes, professores, gestores, escolas, outras instituições de ensino e pesquisa, governo, indústria, associações científicas, entre outros.

Além disto, a formação em Computação é fator estratégico para todos os países, em particular para o Brasil. É importante salientar que devemos primar pela qualidade do ensino em todos os níveis da cadeia de formação. Entendemos que a Computação deva ser ensinada desde o Ensino Fundamental, a exemplo de outras ciências como Física, Matemática, Química e Biologia. Esses são pontos muito importantes para que no futuro tenhamos cidadãos qualificados para enfrentar os desafios do mundo.

Desta forma, considerando que:

1) A sociedade mundial está em constante mudança e estas mudanças acontecem cada vez mais rapidamente, tornando os desafios e problemas cada vez mais complexos;

2) Vivemos em tempos em que a criatividade do homem faz a diferença; a nova economia mundial não se baseia apenas em recursos naturais, mas em conhecimento, fluxos de informação e as habilidades de usá-los;

---

[4] http://www.codeclubbrasil.org

[5] http://www.programae.org.br

3) Os jovens têm experiência e familiaridade na interação e no consumo de novas tecnologias, mas têm pouca experiência em criá-las e expressarem-se com novas tecnologias;

4) A Computação não oferece apenas artefatos de softwares e hardware, mas fundamentalmente uma maneira diferenciada de raciocinar e, compreender e resolver problemas. Todas as pessoas, independentemente da área de formação, se beneficiam ao pensar computacionalmente e ao solucionar problemas através da análise de uma quantidade massiva de dados ou fazer questionamentos que nunca foram cogitados ou ousados devido a sua escalabilidade;

5) A adoção de noções de Computação na Educação Básica é uma preocupação em diversos países, onde a implantação ocorre de forma rigorosa e possui benefícios educacionais (habilidades de reflexão e solução de problemas, compreensão que o mundo está impregnado com a tecnologia digital) e econômicos (alta demanda de profissionais com boa formação);

6) O ensino de Computação na Educação Básica beneficia o desenvolvimento de habilidades e competências essenciais para a vida moderna, independente na área em que atuará. Vale salientar que esta proposta coaduna com ações em diversos países como Alemanha, Argentina, Austrália, Coréia do Sul, Escócia, França, Estados Unidos da América, Finlândia, Grécia, Índia, Israel, Japão, Nova Zelândia, Reino Unido e outros já possuem conceitos Computação em seus currículos;

7) Há abundante evidência científica que comprova que crianças e adolescentes que aprendem a resolver problemas de maneira computacional, melhoram seu desempenho em outras áreas disciplinares, entre elas a Matemática e Línguas;

8) As habilidades trabalhadas na Computação podem ser usadas em diversas áreas e são conhecimentos e técnicas importantes para aumentar as chances de acelerar o desenvolvimento do país, mantendo sua competitividade, apoiar a descoberta científica em outras áreas e potencializar suas capacidades de inovar e criar novas tecnologias;

9) Existe um corpo de pesquisas e iniciativas sólidas no Brasil que promovem a introdução ao Pensamento Computacional na Educação Básica, em especial nos anos finais do Ensino Fundamental e no Ensino Médio. Iniciativas estas que incluem desde abordagens de Computação desconectada, ensino de robótica, ensino de programação, olimpíadas de programação e outros;

10) Trabalhar com conceitos de Computação e Pensamento Computacional não exige necessariamente a presença do computador e pode ser efetivado através de modelos e materiais alternativos que remetem aos conceitos e processos cognitivos envolvidos;

11) O curso de Licenciatura em Computação está amplamente disseminado em diversas regiões do país e busca formar profissionais para atender a demanda de trabalhar com conceitos de Computação e Pensamento Computacional na Educação Básica.

Assim, é fundamental introduzir conhecimentos de Computação e do Pensamento Computacional aos estudantes da Educação Básica em um status de importância similar a de disciplinas tradicionais como a Matemática, Química, Física e a Biologia.

**REFERÊNCIAS**

ANDRADE, D. *et al.* Proposta de Atividades para o Desenvolvimento do Pensamento Computacional no Ensino Fundamental. Anais do Workshop de Informática na Escola, Congresso Brasileiro de Informática na Educação, 2013.

BALANSKAT, A.; ENGELHARDT, K. Computing our future: Computer programming and coding - Priorities, school curricula and initiatives across Europe. 2014. Disponível em: <http://www.eun.org/c/document_library/get_file?uuid=521cb928-6ec4-4a86-b522-9d8fd5cf60ce&groupId=43887>. Acesso em 1 dez. 2015.

BARCELOS, T. S.; SILVEIRA, I. F. Pensamento Computacional e Educação Matemática: Relações para o Ensino de Computação na Educação Básica. In XX Workshop sobre Educação em Computação, 2012.

CFE Argentina. Resolución CFE No 263/15. Disponível em: <http://www.me.gov.ar/consejo/resoluciones/res15/263-15.pdf>. Acesso em 15 out. 2015.

CHOI, J.; AN, S.; LEE, Y. Computing Education in Korea—Current Issues and Endeavors. ACM Transactions on Computing Education, v. 15, n. 2, p. 1–22, 2015.

CLEMENTS, D. H. Computers in Early Childhood Mathematics. Contemporary issues in early childhood, v. 3, n. 2, p. 160–181, 2002.

CODE.ORG. Where computer science counts. Code.org. Disponível em: <https://code.org/action>. Acesso em 4 dez. 2015.

COMPUTING AT SCHOOL. Computing Progression Pathways. Disponível em: <http://community.computingatschool.org.uk/files/5095/original.pdf>. Acesso em 09 dez. 2015.

CSTA Computational Thinking Task Force, Computational Think Flyer. Disponível em: http://csta.acm.org/Curriculum/sub/CurrFiles/CompThinkingFlyer.pdf, Acesso em abr. 2015.

CSTA. K–12 Computer Science Standards - Revised 2011 - The CSTA Standards Task Force. [s.l.]: Association for Computing Machinery, 2011.

DAVIS, J. Australia forgets that code is cultural: replaces History and Geography with Computer Science. Disponível em: <http://thesocietypages.org/cyborgology/2015/10/08/australia-forgets-that-code-is-cultural-replaces-history-and-geography-with-computer-science/>. Acesso em 18 nov 15.

DET - The Department of Education and Training. Taking action now to revitalise STEM study in schools. Disponível em: https://ministers.education.gov.au/pyne/ taking-action-now-revitalise-stem-study-schools. Publicado em 18 set. 2015. Acesso em 13 jan. 2016.

DORAN, K. *et al*. Outreach for improved student performance: a game design and development curriculum. In: [s.l.]: ACM Press, 2012, p. 209. Disponível em: <http://dl.acm.org/citation.cfm?doid=2325296.2325348>. Acesso em 4 dez. 2015.

EPA. Einheitliche Prüfungsanforderungen – Informatik. 2004. Disponível em: <http://www.kmk.org/fileadmin/veroeffentlichungen_beschluesse/1989/1989_12_01-EPA-Informatik.pdf>. Acesso em 10 nov. 2015.

FLEURY, A.; NEVEUX, C. H. "Le code informatique à l'école dès septembre". LeJDD.fr. Disponível em: <http://www.lejdd.fr/Societe/Hamon-Le-code-informatiqu-a-l-ecole-des-septembre-675912>. Acesso em 28 nov. 2015.

FRANÇA, R.; AMARAL, H.. Proposta Metodológica de Ensino e Avaliação para o Desenvolvimento do Pensamento Computacional com o Uso do Scratch. Anais do Workshop de Informática na Escola, Congresso Brasileiro de Informática na Educação, 2013.

GONÇALVES, F. Um instrumento para diagnóstico do Pensamento Computacional. Dissertação de Mestrado. Mestrado em Computação Aplicada - Universidade do Vale do Itajaí, 2015.

JOHNSON, P. France to offer programming in elementary school. ITworld. Disponível em: <http://www.itworld.com/article/2696639/application-management/france-to-offer-programming-in-elementary-school.html>. Acesso em 28 nov. 2015.

JONES, S. P. Computing at School. International comparisons. Disponível em: <http://www.computingatschool.org.uk/index.php?id=documents>. Acesso em 28/Fev/15. Microsoft 2011.

MYKKÄNEN, J.; LIUKAS, L. Koodi 2016. 1. ed. Finlândia: Lönnberg Print, 2014. Disponível em: <https://s3-eu-west-1.amazonaws.com/koodi2016/Koodi2016_LR.pdf>. Acesso em 15 mar. 2015.

NOSCHANG, L. F. ; DE JESUS, E. A. ; PELZ, F. ; RAABE, A. L. A. Portugol Studio: Uma IDE para Iniciantes em Programação. Workshop sobre Educação em Informática, 2014, Brasilia. Anais do Congresso Anual da Sociedade Brasileira de Computação. Porto Alegre: SBC, 2014.

PAPERT, S. Mindstorms: children, computers, and powerful ideas. New York: Basic Books, 1980.

PAPERT, S.. A máquina das crianças: Repensando a escola na era da Informática. Artes Médicas: Porto Alegre. 2008.

PAPERT, S. Redefining Childhood: The Computer Presence as an Experiment in Developmental Psychology. 8th World Computer Congress: IFIP Congress. 1980. Disponível em: <http://www.papert.org/articles/RedefiningChildhood.html>. Acesso em 6 dez. 2015.

PROGRAMAÊ. Programaê. Disponível em: <http://programae.org.br>. Acesso em 14 dez. 2015.

RAABE, A. L. A.; VIEIRA, M. V.; SANTANA, A. L. M.; GONCALVES, F. A.; BATHKE, J. Recomendações para Introdução do Pensamento Computacional na Educação Básica. 4º

DesafIE - Workshop de Desafios da Computação Aplicada à Educação, 2015, Recife. Anais do Congresso Anual da Sociedade Brasileira de Computação. Porto Alegre: SBC, 2015.

RIBEIRO, L.; NUNES, D. J.; CRUZ, M. K.; MATOS, E. S. . Computational Thinking: Possibilities and Challenges. 2nd Workshop-School on Theoretical Computer Science, Rio Grande, RS. WEIT, 2013.

SADOSKY, Fundación. CC-2016: Una propuesta para refundar la enseñanza de la computación en las escuelas Argentinas. 2013. Disponível em: <http://www.fundacionsadosky.org.ar/wp-content/uploads/2014/06/cc-2016.pdf>. Acesso em 10 nov. 15.

SBC - Sociedade Brasileira de Computação. Posição da SBC sobre a Base Nacional Comum Curricular (BNCC). Disponível em <http:www.sbc.org.br>. Acesso em: 15 jan. 2016.

SCC. Surrey Chambers of Commerce. Disponível em: <http://www.surrey-chambers.co.uk/images/qualifchart.gif>. Acesso em 20 ago 2015.

SUPERGEEKS. SuperGeeks. Disponível em: <http://supergeeks.com.br/>. Acesso em 14 dez. 2015.

THE ROYAL SOCIETY. Shut down or restart? The way forward for computing in UK schools. Technology, 2012.

TIC EDUCAÇÃO 2014 - Pesquisa sobre o uso das tecnologias da informação e comunicação nas escolas brasileiras. São Paulo: Comitê Gestor da Internet no Brasil, 2015. Disponível em: http://www.cetic.br/media/docs/publicacoes/2/TIC_Educacao_2014_livro_ eletronico.pdf. Acesso em 13 jan. 216.

TJL. Teaching Jobs Londo. Disponível em: < https://www.youtube.com/watch?v=aDDMEa9E8Z4>. Acesso em 20 ago 2015.

USCONGRESS, 114th. Every Student Succeeds Act. Disponível em: <http://edworkforce.house.gov/uploadedfiles/every_student_succeeds_act_-_conference_report.pdf>.

VALENTE, J. A.. O Professor no Ambiente Logo: Formação e Atuação. 1ª ed. Campinas, NIED Unicamp, 1996.

VIEL, F.; RAABE, A. L. A.; ZEFERINO, C. A. Introdução a Programação e à Implementação de Processadores por Estudantes do Ensino Médio. Workshop de Informática na Escola, 2014, Dourados. Anais do III Congresso Brasileiro de Informática na Educação. Porto Alegre: SBC, 2014.

WING, J. Computacional thinking. Communications of ACM, v. 49, n. 3, p. 33-36, 2006.

WEINBERG, M. Voando para o Futuro. Veja, n. 2431, 2015. (91). Disponível em: <http://veja.abril.com.br/acervo/home.aspx>. Acesso em 30 jun. 2015.

WING, J. M. Computational Thinking Benefits Society. Social Issues in Computing, 2014. Disponível em: <http://socialissues.cs.toronto.edu/2014/01/computational-thinking/>. Acesso em 24 nov. 2015